\documentclass[conference]{IEEEtran}
\pdfoutput=1
\usepackage[utf8]{inputenc}
\usepackage[english]{babel}
\usepackage[T1]{fontenc}
\usepackage{amsmath}
\usepackage{hyperref}
\usepackage{amsmath,amssymb,amsfonts}
\usepackage{algorithmic}
\usepackage{graphicx}
\usepackage{textcomp}
\usepackage{xcolor}
\usepackage[utf8]{inputenc}
\usepackage{subcaption}
\usepackage{url}
\usepackage{braket}
\usepackage[capitalise]{cleveref}
\usepackage[numbers]{natbib}

\ifCLASSINFOpdf
\else
\fi
\begin{document}

\title{Extending the Q-score to an Application-level Quantum Metric Framework}

\author{\IEEEauthorblockN{Ward van der Schoot}
\IEEEauthorblockA{The Netherlands Organisation of\\
of Applied Scientific Research (TNO)\\
Applied Cryptography and Quantum Algorithms\\
The Hague, The Netherlands\\
Email: ward.vanderschoot@tno.nl}
\and
\IEEEauthorblockN{Robert Wezeman}
\IEEEauthorblockA{The Netherlands Organisation of\\
of Applied Scientific Research (TNO)\\
Applied Cryptography and Quantum Algorithms\\
The Hague, The Netherlands\\
Email: robert.wezeman@tno.nl}
\and
\IEEEauthorblockN{Niels M. P. Neumann}
\IEEEauthorblockA{The Netherlands Organisation of\\
of Applied Scientific Research (TNO)\\
Applied Cryptography and Quantum Algorithms\\
The Hague, The Netherlands\\
Email: niels.neumann@tno.nl}
\and
\IEEEauthorblockN{Frank Phillipson}
\IEEEauthorblockA{The Netherlands Organisation of\\
of Applied Scientific Research (TNO)\\
Applied Cryptography and Quantum Algorithms\\
The Hague, The Netherlands\\
Email: frank.phillipson@tno.nl}
\and
\IEEEauthorblockN{Rob Kooij}
\IEEEauthorblockA{The Netherlands Organisation of\\
of Applied Scientific Research (TNO)\\
Cyber Security Technologies\\
The Hague, The Netherlands\\
Email: rob.kooij@tno.nl}}

\author{\IEEEauthorblockN{Ward van der Schoot\IEEEauthorrefmark{1}\IEEEauthorrefmark{3}, Robert Wezeman\IEEEauthorrefmark{1}, Niels Neumann\IEEEauthorrefmark{1}, Frank Phillipson\IEEEauthorrefmark{1}\IEEEauthorrefmark{2}} and Rob Kooij\IEEEauthorrefmark{1}\IEEEauthorrefmark{4}\\
\IEEEauthorblockA{\IEEEauthorrefmark{1} The Netherlands Organisation for Applied Scientific Research,\\
Anna van Buerenplein 1, 2595DA, The Hague, The Netherlands}
\IEEEauthorblockA{\IEEEauthorrefmark{2} Maastricht University, School of Business and Economics,\\
P.O. Box 616, 6200 MD Maastricht, The Netherlands}
\IEEEauthorblockA{\IEEEauthorrefmark{3} Email: ward.vanderschoot@tno.nl}
\IEEEauthorblockA{\IEEEauthorrefmark{4} Faculty of Electrical Engineering, Mathematics and Computer Science, Delft University of Technology, Delft, The Netherlands}}

\maketitle

\begin{abstract}
Evaluating the performance of quantum devices is an important step towards scaling quantum devices and eventually using them in practice. 
The great number of available quantum metrics and the different hardware technologies used to develop quantum computers complicate this evaluation. 
In addition, different computational paradigms implement quantum operations in different ways. 
A prominent quantum metric is given by the Q-score metric of Atos.
This metric was originally introduced as a standalone way to benchmark devices using the Max-Cut problem.
In this work, we show that the Q-score defines a framework of quantum metrics, which allows benchmarking using different problems, user settings and solvers.
To showcase the applicability of the framework, we showcase a second Q-score in this framework, called the Q-score Max-Clique.
This yields, to our knowledge, the first application-level metric capable of natively comparing three different paradigms of quantum computing.
This metric is evaluated on these computational quantum paradigms -- quantum annealing, gate-based quantum computing, and photonic quantum computing -- and the results are compared to those obtained by classical solvers.
\end{abstract}

\begin{IEEEkeywords}
Quantum metrics, Q-score, Application-level benchmarking, Max-Clique
\end{IEEEkeywords}

\IEEEpeerreviewmaketitle

\section{Introduction}
Quantum computers have been under constant development over the last few years and both industry and academia have made tremendous efforts to build ever more powerful quantum computers.
The efforts are however diverse, as there is no consensus on the best approach to building qubits, the fundamental computational units of a quantum computer.
So far, qubits have been based on various physical phenomena, such as superconductivity~\cite{Manucharyan:2009,Houck2009,Barends:2013}, trapped-ions~\cite{Cirac:1995} and photons~\cite{Politi2009}.
Moreover, no consensus exists on the best paradigm for quantum computing. 
Three well-known approaches are quantum annealing, gate-based quantum computing and photonic quantum computing. 
Each hardware technology and even each computational paradigm has its own pros and cons that make it suitable for specific applications. 

While comparing quantum devices based on the same technology is already hard, comparing quantum devices based on different technologies is a research field on its own. 
\textit{Quantum metrics} provide a first step towards quantifying the performance of quantum devices in a unified manner. 
However, just as with quantum computers, a single best metric does not exist. 
An overview of some of the leading metrics, as well as a classification of such metrics, is given by \cite{van_der_Schoot_2023}.
Instead, there exist multiple metrics and each has its own benefits and limitations. 

For a long time, quantum devices were compared based on physical characteristics, such as the number of qubits, global and single qubit coherence times and gate fidelities, for instance by means of randomised benchmarking~\cite{Emerson2005,Danker2009}.
These are all examples of \textit{component-level metrics}, which measure the performance of specific components of a quantum device.
While this gives valuable insights, they do not necessarily say much about the performance of a quantum system as a whole.

Later, devices were benchmarked with \textit{system-level metrics}, which consider the performance of the system as a whole. 
A famous example is given by the quantum volume metric~\cite{quantumvolume}, which measures performance by evaluating which circuits a quantum device can faithfully run.
Complementary to the quantum volume metric, there is the CLOPS metric, which measures the time required to implement these circuits~\cite{CLOPS}. 
Both the quantum volume and the CLOPS metric output a single value, which makes comparing the value relatively easy. 
While such system-level benchmarks manage to consider devices as a whole, their output still says relatively little about their capabilities of solving practical problems.
Moreover, these metrics mainly focus on gate-based quantum devices.

These issues are addressed by \textit{application-level metrics}, which measure the performance of quantum devices on quantum algorithms and practical problems. 
There are various flavours of application-level metrics. 
Some consist of a suite of metrics considering different problems, such as the QED-C suite~\cite{Lubinski:2021_arxiv}, SupermarQ~\cite{tomesh2022supermarq} and QPack~\cite{mesman2022qpack}.
Others focus on a specific problem, such as the Q-score~\cite{Qscore} and Quantum Linpack~\cite{Dong_2021}.
All of these metrics originally focus on gate-based devices.
As gate-based quantum computers have the promise to be universal~\cite{Aharonov:2003,Chow:2012}, this is not unreasonable. 
However, other quantum computational paradigms such as photonic quantum computing and quantum annealing are also relevant and useful for specific problems. 

This issue can be solved by extending the Q-score metric.
The metric was originally defined as a stand-alone number considering the real-world Max-Cut problem to benchmark devices.
The main idea behind the Q-score metric is that a quantum routine usually is part of a larger computational workflow, and benchmarks should therefore consider the whole computational pipeline instead of only the quantum part. 
For this reason, it can be defined in a computing-agnostic way, because of which it can be used to benchmark any machine that can solve the Max-Cut problem. 
Earlier work shows how the Q-score can be extended to benchmark quantum annealers and classical solvers~\cite{Schoot2022EvaluatingQscore} as well.

In this work, we show that the original formulation of the Q-score can be extended to form a quantum metrics framework.
This is due to the fact that there are various parameters in the metric which can be freely chosen, most notably the optimisation problem.
We showcase the strength of this degree of freedom by defining a second Q-score, called the Q-score Max-Clique.
Our Max-Clique extension is specifically designed to yield a metric that can benchmark quantum annealers, gate-based and photonic quantum devices, as well as classical devices. 
This yields, to the best of our knowledge, the first metric capable of making a native comparison between these three paradigms. 
The metric is used to benchmark these three computational quantum paradigms, as well as two solvers based on classical computing. 
To our knowledge, this is the first time performance of these four computational paradigms is natively compared. 

The structure of this paper is as follows: In~\cref{sec:background} background information is provided on quantum computing for the different technology paradigms and earlier Q-score work. 
\cref{sec:qscore-max-clique} explains how the Q-score can be adapted to define a framework, and details a second member of this framework called the Q-score Max-Clique metric. Afterwards, \cref{sec:results} presents the experiments and subsequent results. This work concludes with a discussion of the results and directions for further research in~\cref{sec:discussion}.

\section{Background}\label{sec:background}
\subsection{Quantum Computing}
Quantum computers exploit quantum mechanical principles to perform operations and solve problems.
Contrary to classical states, quantum states do not have to be in a single computational basis state. 
Instead, a quantum state can be any complex linear combination of multiple computational basis states, a so-called superposition. 
Upon measurement, only a single computational basis state is found and the probability depends on the amplitude of that state right before the measurement.
In addition, quantum states exhibit correlations beyond what is possible classically. 
This strong correlation between quantum states is called entanglement. 

No consensus exists on how quantum computers should best exploit these principles.
Some approaches can perform general computations but are harder to build or scale, while other techniques might be easier to implement but are limited in the type of problems they can handle. 
The next three sections briefly recap three popular approaches to quantum computing. 

\subsubsection{Quantum annealing}
Quantum annealing is based on adiabatic quantum computing~\cite{FGGS:2000}, which is known to be equivalent to standard gate-based quantum computing~\cite{Aharonov:2007}. 
In adiabatic quantum computing qubits are not addressed individually, but, instead, qubits are prepared in some initial state and a certain Hamiltonian is slowly applied to these qubits. 
The Hamiltonian is constructed so that its ground-state encodes the solution of the problem.
If this evolution is slow enough, the state remains in the ground state of the Hamiltonian throughout evolution, yielding the solution to the problem. 
The system is required to remain coherent during the evolution time. 
As this evolution time of the process can be exponential in the size of the problem, there are certain limitations to adiabatic quantum computing.

Quantum annealing tries to solve precisely this problem~\cite{KN:1998}, where the Hamiltonian is applied fast and more often. 
While the system is unlikely to remain in the ground-state, more samples get generated, which can approximate the ground-state reasonably well. In its current form, quantum annealing is non-universal and only suited for optimisation problems.
Programming quantum annealing algorithms requires describing the algorithm as a quadratic unconstrained binary optimisation (QUBO) problem.
Many problems have a QUBO formulation~\cite{Lucas:2014} and have been implemented on available hardware for small problem instances. 

\subsubsection{Gate-based quantum computing}
Gate-based quantum computing comes closest to classical digital computers. Individual qubits or groups of qubits are manipulated by means of quantum gates. By careful manipulation of the right qubits, any unitary operation can be implemented on these qubits.
Because of its similarity to classical computers, any classical computing circuit can also be run on a gate-based quantum device. However, most gate-based devices are limited to single qubit gates and specific two-qubit gates on adjacent qubits. 
While this limits the unitary operations a device can perform, this suffices for approximating any unitary operation to arbitrary precision~\cite{Kitaev1997}.


\subsubsection{Photonic quantum computing}
With photonic hardware, photons are used as qubits. 
Photonic quantum devices manipulate these photons using phase shifters and beamsplitters. 
Phase shifters change the phase of the light, whereas beamsplitters entangle two different light waves. 
Finally, squeezers change the position and the momentum of the light wave. 
This is a non-linear interaction needed for universal quantum computing. 
In theory universal computations are possible on photonic quantum devices. However, a main limitation is that universal computations using photonic devices require probabilistic non-linear interactions. Because of this, there is only a certain probability of successfully running the algorithm. 
The depth of circuits that are performed on photonic quantum devices are inherently limited by the length of the used setup. 
To overcome this, some form of rerouting the photons back into the setup while simultaneously changing the operations and the next path for the photons would be needed. Currently, this has not yet been implemented in practice.

\subsection{Q-score}
In 2021, the company Atos proposed the original Q-score metric with the idea that the state-of-the-art benchmarking of quantum devices was insufficient to assess High-Performance Computing platforms~\cite{Qscore}. They designed the Q-score with the idea of having a metric which is: ``application-centric, hardware-agnostic and scalable to quantum advantage processor sizes and beyond". Other strengths of the Q-score are that it is useful (it tells users what they should know), that it is comprehensive (it covers all relevant attributes)~\cite{BCG}, and that it is easy to understand.

In its original definition, the Q-score measures the largest average problem instance of the Max-Cut problem at which a (quantum) device significantly outperforms a random algorithm. The Max-Cut problem is an NP-hard problem in graph theory which entails finding a partition $P$ of the vertices of a (weighted) graph into two sets, so that the total weight of edges between the two sets is maximal. 
The partition $P$ is called a cut and the total weight between these two sets is called the cost $C(P)$. 
Just like many optimisation problems, the Max-Cut problem has a QUBO formulation. 
In addition, the Max-Cut problem is shown to be equivalent to many other optimisation problems.
The Q-score can thus be seen as a metric that measures the performance of a (quantum) device on general optimisation problems.

To gauge the performance of a certain device on the Max-Cut problem, Atos suggests solving the Max-Cut problem for Erd{\"o}s-R{\'e}nyi graphs $G(N,\frac{1}{2})$~\cite{erdos1959random,erdos1960evolution}, which are graphs with $N$ vertices in which each edge exists with probability $\frac{1}{2}$. 
By running the algorithm on sufficiently many random graph instances, the average cost $C(N)$ of the best cut can be computed. This value is then compared to both the average cost of a random cut $C_{rand}$ on the same graphs and the average cost of the best cut $C_{max}$ of the same graphs. The performance of the obtained solution can then be gauged by evaluating the function $\beta(N)$:
$$\beta(N)=\frac{C(N)-C_{rand}}{C_{max}-C_{rand}}=\frac{C(N)-N^2/8}{0.178N^{3/2}},$$
The expression $C_{rand}=N^2/8$ is based upon a random algorithm which randomly partitions the vertices into two sets of size $\lfloor \frac{N}{2}\rfloor$ and $\lceil \frac{N}{2}\rceil$. As each edge exists with probability $\frac12$, the expected cost of this cut is $\lfloor \frac{N}{2}\rfloor\cdot\lceil \frac{N}{2}\rceil\cdot\frac{1}{2}\approx \frac{N^2}{8}$. 
The expression $C_{max}=N^2/8+0.178N^{3/2}$ follows from theoretical results and a least-squared numerical fit for smaller values of $N$~\cite{Qscore}.

From the definition of $\beta(N)$ it follows that a random approach results in a $\beta(N)$ of roughly zero. In contrast, an algorithm that always finds the best possible cut will give $\beta(N)\approx 1$. The value $\beta(N)$ hence captures how `good' a cut is, with higher $\beta(N)$ values corresponding to better cuts. 
The Q-score of a device is defined as the largest $N$ for which $\beta(N) > \beta^*$, for some threshold value $\beta^*$. 
Atos suggest a threshold value $\beta^*=0.2$ to represent a successful execution of the algorithm. Thus, any algorithm achieving a $\beta(N)$ value higher than $0.2$ successfully solved the problem. 
Other choices might suffice as well, however $\beta^*=0.2$ was chosen so that a noiseless gate-based quantum device running the Quantum Approximate Optimisation Algorithm (QAOA)~\cite{FGG:2014} with $p=1$, obtains an infinite Q-score.

Any device that can solve the Max-Cut problem can be assigned a Q-score. This includes classical devices, gate-based quantum computers and quantum annealing devices. The Q-score hence allows for comparing devices based on different computational paradigms. Note that the Q-score depends not only on the used backend device, but also on the used algorithm, which gives freedom to use different algorithms and also compare them to each other. Naturally, this calls for mentioning the used algorithm to compute the Q-score. 

The original definition of the Q-score allows the user to choose some variables, such as the allowed time for the computation. This impacts the Q-score significantly, as classical devices would be able to solve the problem for any $N$ perfectly, resulting in an infinite Q-score, given a sufficient amount of time. Other examples of user variables to choose are the number of random instances of Erd{\"o}s-R{\'e}nyi graphs and the allowed pre- and post-processing optimisation steps. 
Earlier work~\cite{Schoot2022EvaluatingQscore} imposed a time-constraint of at most $60$ seconds for each instance, a total of $100$ random graph instances per graph size and considered each solver in an out-of-the-box fashion to restrict any pre- or post-processing optimisation steps. 
If a device failed to find an answer within $60$ seconds, a random cut was used for that instance, corresponding to a value of $\beta(N)=0$.
This earlier work showed that, given these constraints, classical backends achieve higher Q-scores ($2{,}300$ and $5{,}800$) than state-of-the-art quantum annealing hardware ($70$ and $140$). 
In turn, the annealing-classical hybrid method offered by D-Wave Systems Inc. achieved a Q-score of $12{,}500$, which hints at gains offered by quantum annealing. 
The work of \cite{Schoot2022EvaluatingQscore} also briefly discusses gate-based devices and their expected lower Q-score, mainly due to their low number of available qubits, and does not discuss photonic devices.

\section{The Q-score Framework and the Q-score Max-Clique}\label{sec:qscore-max-clique}
As mentioned in the previous section, the original Q-score definition leaves a couple degrees of freedom, the most notable ones being the allowed runtime and the amount of optimisation.
Earlier work imposed specific values for these degrees of freedom~\cite{Schoot2022EvaluatingQscore}, which yielded a completely set metric. 
In addition, the original Q-score definition makes specific choices for parameters which are inherently degrees of freedom, the most notable ones being the optimisation problem, the considered data set and the $\beta^*$-value, which measures the required accuracy.
However, these are actually degrees of freedom which should be imposed by the use case.
As (application-level) quantum metrics should optimally measure the performance of devices on applications, these parameters should also remain a degree of freedom in the definition of the metric.

This results in the definition of the \textit{Q-score framework}, in which optimisation problem, data set, $\beta^*$-value, allowed runtime and allowed optimisation are all degrees of freedom which have to be chosen by the end user.
Given a certain use case, the Q-score for this specific use case can then be defined by taking the following steps:
\begin{enumerate}
\item Choose (optimisation) problem that is most relevant to the use case. This problem needs to satisfy three requirements: firstly, it needs to be solvable for increasing problem sizes. Secondly, it needs to have a single value as an output, with a higher value corresponding to a better performance.
Lastly, and most importantly, it needs to have a scalable approximation for the average value of a random and the optimal solution.
\item Choose a data set relevant to the use case on which devices can try to solve the considered problem.
\item Choose the required accuracy for the use case, resulting in a certain $\beta^*$-value.
\item Choose the allowed runtime for solving a single problem instance.
\item Choose the amount of optimisation that is allowed for solving a single problem instance. This amount should directly depend on the amount of (economic) resources that are available for the end user.
\end{enumerate}
The resulting Q-score results in a metric that optimally measures the performance of a device at solving the use case.

It is good to emphasise that the Q-score framework in itself is nothing more than a way of combining the different degrees of freedom to yield a single number output.
In the general definition, the Q-score framework depicts the largest problem size for which a device "on average" "significantly" outperforms a random approach at a certain problem on a certain data set, within the given time limit and allowed optimisation. 
In this definition, on average means averaged over 100 instances within the given data set. 
It should be noted that this number could also be seen as a degree of freedom, but it is not expected that changing this value will result in different metrics.
A device significantly outperforms a random solution, if the cost $C(N)$ for a given problem size $N$ satisfies
$$
\beta(N):=\frac{C(N)-C_{rand}}{C_{max}-C_{rand}}>\beta^*,
$$
where $C_{rand}$ and $C_{max}$ are the costs of a random and optimal solution of the considered problem.

To avoid confusion between different instantiations of the Q-score metric, we suggest to use the standalone name of Q-score only for the framework itself.
To talk about a specific instantiation of the Q-score, the name Q-score accompanied with the optimisation problem should be used.
If the other degrees of freedom are of relevance, these should be mentioned as well.
Thus, the Q-score definition from~\cite{Schoot2022EvaluatingQscore} should be called the Q-score Max-Cut within this framework (on a data set of Erdös-Renyí ($N,\frac12)$-graphs with $\beta^*0.2$, 60 seconds runtime limit and no allowed (extra) optimisation).

\subsection{Q-score Max-Clique}
To showcase the strength of the framework, a second instantiation of the Q-score framework will be given in this section, called the Q-score Max-Clique. 
The idea of this extension is to have a metric which can be be natively implemented on photonic quantum devices, in addition to gate-based quantum devices, quantum annealers and classical devices. 
This allows a fair comparison between all of these paradigms.
The next section will show that this metric can indeed be used to benchmark all four of these paradigms.

The \textit{Max-Clique problem} is an established NP-hard problem in graph theory which aims to find the largest complete subgraph of a random graph of size $N$. A complete subgraph, or clique, is a subgraph in which every two vertices are connected by an edge. 
Just like the Max-Cut problem, the Max-Clique problem has a QUBO-formulation and the problem is hence equivalent to many optimisation problems, in particular to the Max-Cut problem~\cite{QUBO_problems}. The Q-score Max-Clique hence also serves as a metric on how well a device performs at general optimisation problems.

To define a Q-score using the Max-Clique problem, the Max-Clique problem needs to satisfy the three requirements as mentioned above, of which it clearly satisfies the first two.
For the third requirement, a scalable approximation of the average random and optimal cost for solving the Max-Clique problem is required.

The random cost $C_{rand}$ is defined as as the average clique size found by the following random naive approach: start with an empty sub-graph $G'$; Take a random node $v\in G\setminus G'$ and add this to $G'$ if, $G'\cup\{v\}$ is still a clique. If not, return the number of nodes in $G'$ as the maximum clique size found by this random algorithm. 
For general Erd{\"o}s-R{\'e}nyi graphs $G(N,p)$, the probability that this naive algorithm finds a clique of size at least $i$ is 
$$\mathbb{P}[X\geq i] = p^{i(i-1)/2}.$$
This follows as each of the $i(i-1)/2$ edges has to be present and each edge is added to the graph independently with probability $p$. 
From this is follows that 
\begin{align*}\mathbb{P}[X=i] &= \mathbb{P}[X\ge i] - \mathbb{P}[X\ge i+1]\\
&= p^{i(i-1)/2} - p^{i(i+1)/2} = (1-p^i)p^{i(i-1)/2}.\end{align*}
The expected size of the clique found with the naive algorithm then equals
$$\mathbb{E}[X] = \sum_{i=1}^N i\cdot\mathbb{P}[X=i] = \sum_{i=1}^N i\cdot (1-p^i)p^{i(i-1)/2}.$$
For $p=\frac{1}{2}$, this converges quickly to the approximate value given by $C_{rand}=\mathbb{E}[X] = 1.6416325$.

For the average cost $C_{max}$ of the optimal solution, there exist scalable expressions in literature~\cite{matula1976largest}, which say that this value scales as
\begin{align*}
    C_{max} &= 2\log_2(N) - 2\log_2(\log_2(N)) + 2\log_2(e/2) + 1\\
    &= 2\log_2\left(\frac{Ne/2}{\log_2(N)}\right)+1.
\end{align*}
Note that this is an asymptotic result for $N\to \infty$ and might not be a good approximation for smaller $N$.
With these two definitions, it can be concluded that the Max-Clique problem can indeed be used to define a Q-score metric.

In this work, we consider a specific instantiation in which the other degrees of freedom are also set.
Firstly, as a data set, the same type of graphs as the original Q-score: $(N,\frac{1}{2})$-Erd{\"o}s-R{\'e}nyi graphs, and and the cost $C(N)$ is computed as the average clique size found over $100$ random instances of these graphs.
Secondly, a value of $\beta^*=0.2$ is taken again, following~\cite{Qscore}. 
Thirdly, the limit on the computation time is set to $60$ seconds per instance, as in~\cite{Schoot2022EvaluatingQscore}.
If a device does not find an answer within the allocated 60 seconds, a clique size of $C_{rand}$ is assigned, corresponding to a value of $\beta(N)=0$. 
Finally, all algorithms are considered in an out-of-the-box fashion to restrict any additional optimisation steps.

All in all, the Q-score Max-Clique with these degrees of freedom is thus the largest problem size $N$ for which an unoptimised device obtains a cost $C(N)$ within 60 seconds for which 
\begin{align*}
  \beta(N) & =\frac{C(N)-C_{rand}}{C_{max}-C_{rand}} \\
  & = \frac{C(N) - 1.6416325}{2\log_2\left(\frac{Ne/2}{\log_2(N)}\right) - 0.6416325} > 0.2.
\end{align*}

\section{Results}\label{sec:results}
In this section the defined Q-score Max-Clique will be utilised to benchmark the different computational paradigms.
It should be noted that the goal of this is not the benchmark itself, but rather to show that the metric can indeed be used to benchmark all the different platforms in a native way. 
First, a brief overview is given of how the Max-Clique problem can be solved on each of the quantum computing paradigms, as well as the devices for which the Q-score Max-Clique is computed.

\begin{itemize}
    \item \textbf{Quantum Annealing}: The literature contains many QUBO formulations for the Max-Clique problem, such as the one in~\cite{QUBO_problems}. This formulation is taken and then solved using annealing hardware from D-Wave Systems Inc. Both their 2000Q and Advantage systems, with $2041$ qubits and $5627$ qubits, respectively, are considered. 
    
    \item \textbf{Gate-based quantum computing}: The same QUBO formulation can be used to solve Max-Clique instances using the Quantum Approximate Optimisation Algorithm (QAOA)\cite{FahriGoldstoneGutmann:2014}.
    This is a hybrid quantum-classical algorithm in which a certain cost function, depending on measurements of quantum states, is optimised with a classical optimiser. 
    For each optimisation step, a new quantum circuit needs to be run.
    Currently, quantum devices are accessed through the cloud, and resources are shared with other users.
    Because of this, each set of quantum circuits needs to be sent as a job through the entire stack and queue, which is rather inefficient, especially in variational settings, such as QAOA. 
    Solutions are being developed that provide users the ability to avoid latency by executing entire programs in containerised services, such as Qiskit Runtime~\cite{runtime}.
    Still, significant queue times remain and once scheduled, a single problem instance can still take around an hour to execute in practice. The actual computation time QAOA would take with direct access to dedicated hardware is hard to estimate. A methodology to reason about the execution-time of QAOA is developed in~\cite{Weidenfeller2022}. According to their estimates, candidate solutions for Max-Cut graphs with $500$ nodes can be generated in under three minutes, which is still above the $60$ seconds threshold.
    
    An extensive review of QAOA, including its various variants and the challenges related to their applicability, is provided in \cite{BLEKOS20241}.
    
    A simple QAOA implementation is considered using only one layer with SLSQP as optimisation algorithm, see~\cite{github} for the exact implementation. For the experiments, two hardware backends from different vendors are used: the 16 qubit Guadalupe device~\cite{guadalupe16} by IBM, and the 5 qubit Starmon-5 device~\cite{starmon5} by Quantum-Inspire. Additionally, noiseless simulations without the time constraint are performed to upper-bound what real hardware could currently achieve.


    \item \textbf{Photonic quantum computing}: The main component used in photonic quantum computing to solve Max-Clique instances is called Boson sampling~\cite{AA:2013}. 
    Boson sampling can be used to obtain dense subgraphs from a graph~\cite{Arrazola_2018}. By removing the least-dense vertices of such subgraphs, they can be converted into cliques. Such a clique can be made (locally) maximal by adding fully connected vertices to it. It should be noted that the latter two steps are classical steps, making this algorithm a photonic-classical hybrid algorithm. It was shown that the resulting hybrid procedure is able to find the largest cliques with sufficiently high probability~\cite{Banchi2020}. A more extensive overview of using boson sampling for solving Max-Clique problems can be found in~\cite{Haverly2021}. For the experiments, the Ascella QPU backend by Quandela is used~\cite{quandela}. Additionally, noiseless simulations without the time constraint are performed using Strawberry Field's GBS sampler~\cite{photonic_sampler} to upper-bound what real hardware could currently achieve.
\end{itemize}

In addition, the Q-score Max-Clique is also calculated for two classical solvers, namely qbsolv and simulated annealing, both offered by D-Wave Systems Inc. The qbsolv method solves smaller pieces of a QUBO with a tabu search algorithm~\cite{Glover1989,Glover1990} and then combines those solutions to a full solution. 
The simulating annealing method is able to solve a QUBO by searching the solution space and slowly decreasing the probability of accepting worse solutions than the current optimum. Lastly, the Q-score Max-Clique is computed for the hybrid solver provided by D-Wave Systems Inc. This is an algorithm which combines classical and annealing approaches to solve QUBOs. The precise workings of this solver are however unknown, so the solver should be seen as a black-box device, which this may yield a somewhat unfair comparison. However, this is not an issue for this work, as the goal is to show that the Q-score Max-Clique can be used to benchmark all the different platforms, not to perform the comparison itself. This also demonstrates that the Q-score metric can be used to compare with various classical (approximate) algorithms.

\cref{tab:overview} gives the Q-score of all considered backends.

\begin{table}
    \centering
    \caption{Q-scores Max-Clique with a $60$ seconds time limit. The starred entries were obtained using more than 60 seconds.}
    \begin{tabular}[t]{l|r}
        Approach &  Q-score \\
        \hline
        D-Wave Tabu search & 4,900\\
        D-Wave Simulated annealing & 9,100\\
        D-Wave Advantage & 110\\
        D-Wave 2000Q & 70\\
        D-Wave Hybrid solver & 12,500\\
        Starmon-5 (QAOA) & 5* \\
        IBM-Guadalupe (QAOA) & $\geq 5$*\\
        Quandela-Ascella & 3
    \end{tabular}
    \begin{tabular}[t]{l|r}
    Approach & Simulated Q-score \\
    \hline
    QAOA & $\geq16$* \\
    Photonics & $\geq20$*\\
    \end{tabular}
    \label{tab:overview}
\end{table}

\subsection{Experiments}
To compute the Q-score Max-Clique, $100$ random Erd{\"o}s-R{\'e}nyi graphs $G(N,\frac{1}{2})$ are generated for increasing problem sizes $N$ with the NetworkX python library. 
Next, the Max-Clique problem for these graphs is solved by the different solvers in an out-of-the-box fashion with the imposed time constraint of 60 seconds. This time considers the entire computation time starting from either a graph or QUBO representation. 
The only exception is the QAOA algorithm, for which only $10$ graph instances are considered and no time limit is imposed due to long queuing times.
For the simulations of the gate-based and photonic approach, the time constraint is also ignored as the simulation time does not represent the time it would take real hardware to solve the problem. 
After obtaining all results for the graphs, the average cost $C(N)$ and corresponding $\beta(N)$ are computed.
The graph size $N$ is increased until $\beta(N)\leq 0.2$.
For the classical algorithms we use a modest server with an Intel Core i7-7600U CPU with a clock speed of 2.80GHz and 12 GBs of RAM.
All code used to perform our experiments has been made open source and can be found at \cite{github}.

Below, the found results for each of the computational paradigms are discussed in detail.

\begin{itemize}
    \item \textbf{Quantum Annealing}:
    The $\beta$ vs $N$ graph for the D-Wave Advantage device and its previous-generation 2000Q device are shown in~\cref{fig:annealing_hardware}. 
    The experiments for both devices start at $N=5$ and in each new step, increase $N$ by five. For the Advantage system, $\beta= 0.2$ occurs for $N=110$, which yields a Q-score Max-Clique of $110$. 
    For the 2000Q device it is not possible to find a problem embedding for instances larger than approximately $N=70$. Hence, a Q-score Max-Clique of $70$ is found. Interestingly, similar limitations exist for Q-score Max-Cut at the same problem size~\cite{Schoot2022EvaluatingQscore}. It should be noted that \cite{Pelofske} deals with this issue by investigating methods for decomposing larger problem instances for Max-Clique into smaller ones, which can subsequently be solved on the 2000Q device.
    
    For small $N$, the approximation $C_{max}$ is suboptimal, which explains the counter-intuitive behaviour seen in the $\beta$ vs $N$ graph for small $N$. 
    The crossing $\beta=0.2$ is unaffected by this approximation error. 
    For completeness, \cref{fig:annealing_bruteforce} gives the $\beta$ vs $N$ graphs with $C_{max}$ the classically computed largest clique, instead of an approximation value. 
    \begin{figure}
        \centering
        \includegraphics[width=0.55\textwidth]{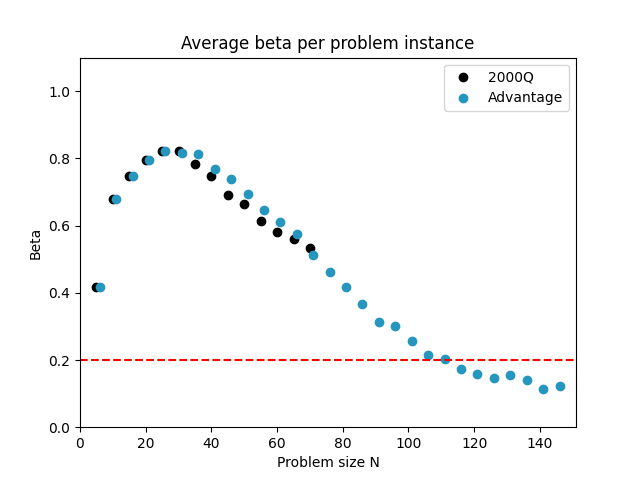}
        \caption{Max-Clique Q-score's $\beta$ vs problem size $N$ for the D-Wave Advantage and 2000Q QPUs. For visibility reasons, the Advantage graph is off-set by $1$ in the x-dimension.}
        \label{fig:annealing_hardware}
    \end{figure}
    \begin{figure}
        \centering
        \includegraphics[width=0.55\textwidth]{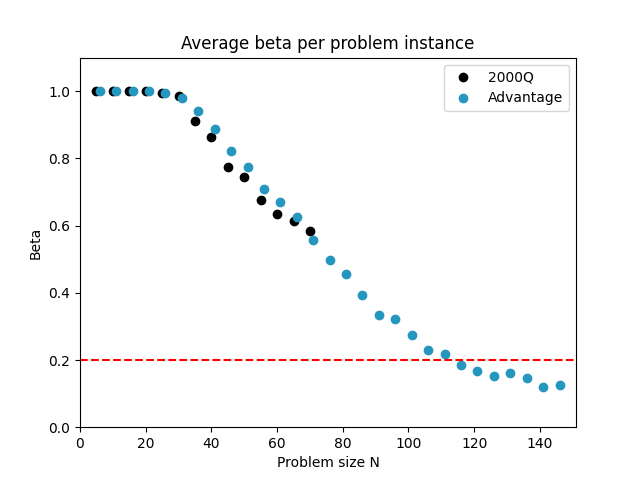}
        \caption{Max-Clique Q-score's $\beta$ vs problem size $N$, with respect to optimal clique solution $C_{max}$, for the D-Wave Advantage and 2000Q QPUs. For visibility reasons, the Advantage graph is off-set by $1$ in the x-dimension.}
        \label{fig:annealing_bruteforce}
    \end{figure}
    
    \item \textbf{Gate-based quantum computing}:
    The Q-score Max-Clique was computed for two hardware backends up to problem sizes $N=5$. On both the Guadalupe and Starmon-5 devices, no solutions were found within the 60 seconds time constraint. This is why it was decided to consider the solvers without the time constraint instead. Under these conditions, perfect clique solutions were found for all tested problem instances. Hence, both devices achieve $\beta(N)=1$ for all $N\le 5$.
    The Starmon-5 device has five qubits, so no larger problem instances can be solved on this device. The Guadalupe device has sixteen qubits, but time-out errors in the experiments prevented computing the Max-Clique instances for $N\geq 6$. 
    
    Noiseless simulations without the time constraint have been performed up to problem sizes $N=16$ to upper-bound the achievable $\beta$ on the Guadalupe device. \cref{fig:simulations} shows the $\beta$ vs $N$ graph.
    From the graph, we conclude that a Q-score Max-Clique of $16$ is in principle attainable using the given QAOA setup.
        \begin{figure}
        \centering
        \includegraphics[width=0.55\textwidth]{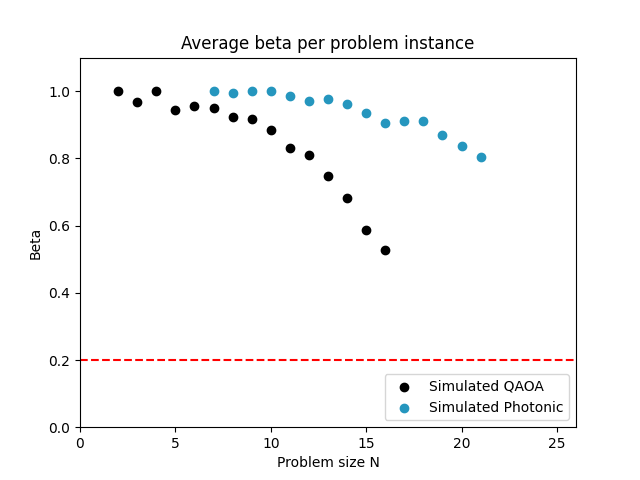}
        \caption{
        Max-Clique Q-score's $\beta$ vs problem size $N$, with respect to optimal clique solution as $C_{max}$, for noiseless simulations of gate-based quantum computing and photonic quantum computing.}
        \label{fig:simulations}
    \end{figure}

\item \textbf{Photonic quantum computing}:
    The Q-score Max-Cliqe was computed for the Ascella QPU offered by Quandela. The hybrid workflow as mentioned above was used, without the use of the classical clique extending step. In this way, the performance of the quantum processor was measured as purely as possible. The experiments gave good results for size $N=3$, with a $\beta(3)=0.71$ as a result. Unfortunately, the experiments for larger problem sizes yielded circuits which were too large for the current hardware offered by Quandela. Because of this, the Q-score obtained by this quantum device equals 3. 
    
    Noiseless simulations without the time constraint have been performed up to problem sizes $N=20$ to upper-bound the achievable $\beta$ on photonic devices. The simulations were run up till $N=20$, as larger problem sizes yielded intractable simulation times. \cref{fig:simulations} shows the $\beta$ vs $N$ graph.
    From the graph, we conclude that a Q-score Max-Clique of $20$ is in principle attainable using a large enough photonic device.

\item \textbf{Classical and hybrid}:
    Lastly, \cref{fig:classical} shows the $\beta$ vs $N$ graph for D-Wave's classical and hybrid solvers.
    For the classical solvers, the simulations start for $N=100$ and increase in size with steps of $100$. 
    The computation is terminated when the average time for graph sizes exceeds $100$ seconds on our server to ensure that the region of 0 to $60$ seconds is captured completely.
    For the hybrid solver, $\beta$ is calculated for problem instances starting from $N=1{,}500$ up to $N=12{,}500$ with steps of $500$ and a time constraint of $60$ seconds. 
    Due to the relative high cost of resources of the D-Wave hybrid solver, only a single problem instance is considered per problem size. 
    
    Each of these solvers is able to maintain a $\beta>0.2$ for large problem instances $N$. Therefore, the Q-score Max-Clique is determined by the imposed 60 seconds time constraint. \cref{fig:classical_times} shows the average time taken to solve a problem instance. For the hybrid solver, the minimum needed computation time is hard-coded.
    From these results, we report a Q-score Max-Clique of $9{,}100$ and $4{,}900$ for the D-Wave implementations of simulated annealing and tabu search respectively. The hybrid solver has a Q-score Max-Clique of $12{,}500$ because of the hard-coded computation time constraint. 
    
    \begin{figure}
        \centering
        \includegraphics[width=0.55\textwidth]{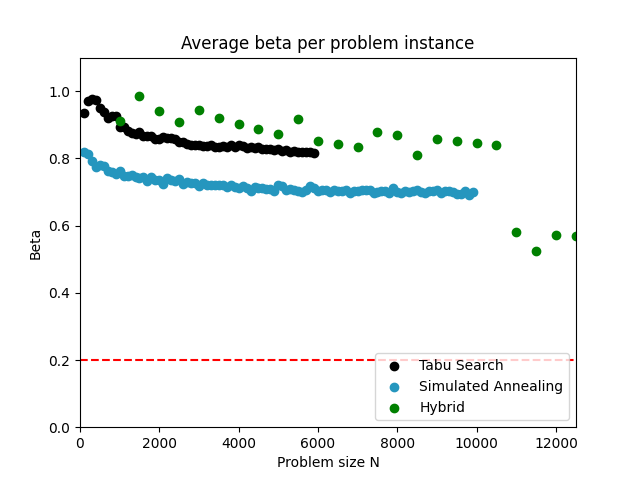}
        \caption{Max-Clique Q-score's $\beta$ vs problem size $N$ for the D-Wave classical tabu search solver (qbsolv), D-Wave Simulated Annealing and their hybrid solver.}
        \label{fig:classical}
    \end{figure}
    \begin{figure}
        \centering
        \includegraphics[width=0.55\textwidth]{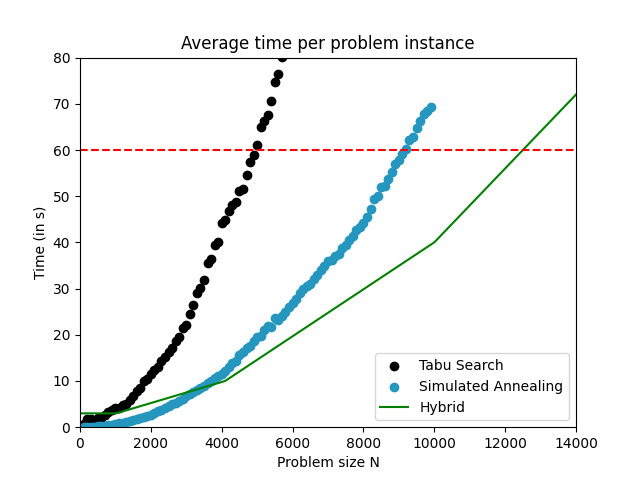}
        \caption{Average time taken per problem instance $N$ for the D-Wave classical tabu search solver (qbsolv), D-Wave Simulated Annealing and the hybrid solver.}
        \label{fig:classical_times}
    \end{figure}
\end{itemize}

\section{Conclusion and Discussion}\label{sec:discussion}
In this work, the Q-score as originally introduced by Atos is extended to a framework definition. 
First, it is noted that the original Q-score definition and subsequent work make many arbitrary choices, which should actually be left as degrees of freedom for the user.
By stripping the original Q-score of these arbitrary choices, the Q-score framework can be defined.
In this framework, optimisation problem, data set, $\beta^*$-value, allowed runtime and allowed optimisation steps are left as degrees of freedom.
By choosing these degrees of freedom to optimally match the use case at hand, a suitable metric can be designed for each use case.
The Q-score framework then captures a score defined as the largest problem size for which a device on average significantly outperforms a random approach.

To showcase the applicability of the Q-score framework, this work also defines the Q-score Max-Clique.
Our choice for Max-Clique is motivated by the existence of a clique search algorithm for photonic hardware.
This allowed us to define the first quantum metric which can be natively implemented on three quantum paradigms, as well as classical computers.
This metric is instantiated with specific degrees of freedom, after which it is applied to solvers from all four paradigms.
This evaluation shows that it is indeed possible to use the Q-score Max-Clique to benchmark all paradigms.

It is important to note that there are other ways of defining quantum metrics that can compare all four paradigms. 
As an example, the BACQ initiative focuses on the original Q-score metric to benchmark photonic devices as well~\cite{barbaresco2024bacq}.
This can be done, as the Max-Cut problem can be rewritten as a problem that is solvable on photonic quantum devices, such as the Max-Clique problem.
Both methods are valid ways of yielding a metric that is capable if benchmarking solvers of all paradigms, with neither being better than the other.
One showcases the value of the Q-score as a framework, while the other showcases the strength of the original Q-score Max-Cut metric itself.

The strength of the Q-score framework comes from the fact that it contains many degrees of freedom that can be optimally matched to different use cases. 
It could be argued that this is also a weakness of the Q-score: users should be very specific on how the metric is defined, which might complicate a potential comparison between different works.
Hardware vendors, in particular, have an incentive to select the degrees of freedom in a way that suit the performance of their hardware.
This makes the Q-score framework less suitable to form "the" application-level metric that should be used in any situation.
However, that is also not the goal of the framework: instead, it yields a method that has gives valuable insights for any use case. 

In the definition of the Q-score Max-Clique, various degrees of freedom were chosen in quite an arbitrary way.
Of course, different choices could be made for the various degrees of freedom, some of which will be discussed here.
Firstly, one could argue that our choice for random algorithm is too naive resulting in a constant $C_{rand}$ value, which might seem strange for large problem instances. 
A different, yet naive, algorithm would be to start from a random node and grow it to a clique by iterating over all possible vertices. 
In the end, different random algorithms will result in different $\beta$ vs $N$ graphs and hence different Q-scores. 
However, it is expected that this does not significantly change the relative performance between devices, and hence does not change the relative Q-scores.


Secondly, the runtime limit was chosen as 60 seconds. 
Although other choices would have been reasonable as well, this was decided following earlier work~\cite{Schoot2022EvaluatingQscore}. 
Unfortunately, this time limit was too high for the photonic and gate-based quantums solvers, mainly due to queueing times. To showcase results for these solvers as well, it was decided to perform the benchmark for these paradigms without the time constraint. 
It was decided to keep the time limit for the other solvers, as especially the classical solvers would have otherwise obtained unrealistically high Q-scores due to their brute-force abilities.
Naturally, this makes the comparison using this benchmark somewhat unfair.
However, as mentioned before, this is not an issue for this work, as the goal of this work is not to perform a fair comparison, but rather the showcase the applicability of the Q-score Max-Clique.
Once the different paradigms have matured, and queueing times have decreased, it is expected that using the same time limit throughout will result in a fair benchmark.

In this work, the hybrid solver of D-Wave obtained the highest Q-score Max-Clique of $12{,}500$ of all considered solvers. The differences with other classical solvers are however not yet significant enough to hint towards a quantum advantage.
The hybrid solver forms a black box, which hinders attributing the better performance solely to the quantum routines.
Unfortunately, D-Wave does not publish any information regarding these precise workings.
What is known, however, is that the hybrid solver uses a server stronger than our local solver for its classical routines. 
Naturally, this yields an unfair comparison, and it is in particular also expected that the classical solvers will obtain higher scores if they were to be implemented using stronger servers.
All in all, it can be concluded that the performed benchmark should not be used to compare the solvability of the hybrid solver to the classical solvers.
Again, this is not an issue for this work.
For future practical use of the Q-score Max-Clique, it is expected that both solvers will be benchmarked using similar server capacity.
In addition, the $\beta$ vs $N$ graph for the hybrid solver shows a seemingly inexplicable drop at around $N=11000$. It seems that there is something in the black box workings of the hybrid solver that results in problems of size larger than $N=10000$ being relatively more difficult to solve for the hybrid solver.

Quantum annealing has the largest Q-score Max-Clique of the different quantum computing approaches, with $110$ for the Advantage system as its highest value. 
This is not a surprise as it has more qubits and hence allows for larger problems to be implemented than the considered gate-based and photonic hardware. 
In turn, gate-based and photonic hardware obtain similar Q-scores, with gate-based devices slightly outperforming their photonic counterpart.
It should however be mentioned that the gate-based devices did not manage to find a solution within the proposed time limit, while the photonic device did.
As QAOA is effectively a smart guessing algorithm, giving it large amounts of time means it would essentially always find the optimal solution for small problem instances.
Future experiments with larger photonic and gate-based devices will have to show how these paradigms actually compare on the Q-score Max-Clique.
Lastly, all classical solvers currently outperform the state-of-the-art quantum computers for each considered technology, as was originally expected.

Comparing Q-score Max-Clique with earlier results for Max-Cut~\cite{Schoot2022EvaluatingQscore}, shows that the Q-score for the Advantage system is slightly lower ($110$ versus $140$), while the 2000Q device finds the same score of 70 for both Q-scores.
On the contrary, the classical approach performs better for the Max-Clique problem, with scores of $2{,}300$ vs $4{,}900$ and $5{,}800$ vs $9{,}100$ for the D-Wave implementations of tabu search and simulated annealing respectively. However, upon closer inspection that does not tell the whole story. Specifically, for the Q-score Max-Cut, the classical and hybrid solvers find solutions of perfect quality $\beta=1$ throughout, while for Q-score Max-Clique, the same solvers show both a variety and decrease in the solution quality parameter $\beta$. So, while the absolute Q-scores Max-Clique are higher, the Q-score Max-Cut finds higher quality solutions throughout.
This hints towards the added value of a suite of Q-scores for different optimisation problems and parameter settings, for example using different $\beta^*$-values.
In this way, the best hardware for each optimisation problem and application can be found.

For future work, it would naturally be most interesting to consider other instantiations of the Q-score framework. 
This should consider both other problems, as well as changing the other degrees of freedom.
In addition, it would be interesting to consider how evaluations using the different Q-scores compare, and how these different Q-scores compare to other application-level metrics.

Regarding future work for the Q-score Max-Clique, it would be interesting to extend the QAOA results to more hardware backends and larger problem instances. For a fair comparison, additional research is needed on the execution time of QAOA.
Alternatively, it would also be interesting to consider different gate-based algorithm approaches. 
For example, one could also solve the Max-Clique problem using a Grover's search~\cite{haverly2021grov} or a variational quantum eigensolver~\cite{Tilly_2022}.

\bibliographystyle{IEEEtran}
\bibliography{main.bib}

\begin{thebibliography}{10}
\providecommand{\url}[1]{#1}
\csname url@samestyle\endcsname
\providecommand{\newblock}{\relax}
\providecommand{\bibinfo}[2]{#2}
\providecommand{\BIBentrySTDinterwordspacing}{\spaceskip=0pt\relax}
\providecommand{\BIBentryALTinterwordstretchfactor}{4}
\providecommand{\BIBentryALTinterwordspacing}{\spaceskip=\fontdimen2\font plus
\BIBentryALTinterwordstretchfactor\fontdimen3\font minus
  \fontdimen4\font\relax}
\providecommand{\BIBforeignlanguage}[2]{{%
\expandafter\ifx\csname l@#1\endcsname\relax
\typeout{** WARNING: IEEEtran.bst: No hyphenation pattern has been}%
\typeout{** loaded for the language `#1'. Using the pattern for}%
\typeout{** the default language instead.}%
\else
\language=\csname l@#1\endcsname
\fi
#2}}
\providecommand{\BIBdecl}{\relax}
\BIBdecl

\bibitem{Manucharyan:2009}
\BIBentryALTinterwordspacing
V.~E. Manucharyan, J.~Koch, L.~I. Glazman, and M.~H. Devoret, ``Fluxonium:
  Single {C}ooper-pair circuit free of charge offsets,'' \emph{Science}, vol.
  326, no. 5949, pp. 113--116, 2009. [Online]. Available:
  \url{https://www.science.org/doi/abs/10.1126/science.1175552}
\BIBentrySTDinterwordspacing

\bibitem{Houck2009}
\BIBentryALTinterwordspacing
A.~A. Houck, J.~Koch, M.~H. Devoret, S.~M. Girvin, and R.~J. Schoelkopf, ``Life
  after charge noise: recent results with transmon qubits,'' \emph{Quantum
  Information Processing}, vol.~8, no. 2-3, pp. 105--115, Feb. 2009. [Online].
  Available: \url{https://doi.org/10.1007/s11128-009-0100-6}
\BIBentrySTDinterwordspacing

\bibitem{Barends:2013}
\BIBentryALTinterwordspacing
R.~Barends, J.~Kelly, A.~Megrant, D.~Sank, E.~Jeffrey, Y.~Chen, Y.~Yin,
  B.~Chiaro, J.~Mutus, C.~Neill, P.~O'Malley, P.~Roushan, J.~Wenner, T.~C.
  White, A.~N. Cleland, and J.~M. Martinis, ``Coherent {J}osephson qubit
  suitable for scalable quantum integrated circuits,'' \emph{Phys. Rev. Lett.},
  vol. 111, p. 080502, Aug 2013. [Online]. Available:
  \url{https://link.aps.org/doi/10.1103/PhysRevLett.111.080502}
\BIBentrySTDinterwordspacing

\bibitem{Cirac:1995}
\BIBentryALTinterwordspacing
J.~I. Cirac and P.~Zoller, ``Quantum computations with cold trapped ions,''
  \emph{Phys. Rev. Lett.}, vol.~74, pp. 4091--4094, May 1995. [Online].
  Available: \url{https://link.aps.org/doi/10.1103/PhysRevLett.74.4091}
\BIBentrySTDinterwordspacing

\bibitem{Politi2009}
\BIBentryALTinterwordspacing
A.~Politi, J.~Matthews, M.~Thompson, and J.~O{\textquotesingle}Brien,
  ``Integrated quantum photonics,'' \emph{{IEEE} Journal of Selected Topics in
  Quantum Electronics}, vol.~15, no.~6, pp. 1673--1684, 2009. [Online].
  Available: \url{https://doi.org/10.1109/jstqe.2009.2026060}
\BIBentrySTDinterwordspacing

\bibitem{van_der_Schoot_2023}
\BIBentryALTinterwordspacing
W.~van~der Schoot, R.~Wezeman, P.~T. Eendebak, N.~M.~P. Neumann, and
  F.~Phillipson, ``Evaluating three levels of quantum metrics on
  quantum-inspire hardware,'' \emph{Quantum Information Processing}, vol.~22,
  no.~12, Dec. 2023. [Online]. Available:
  \url{http://dx.doi.org/10.1007/s11128-023-04184-x}
\BIBentrySTDinterwordspacing

\bibitem{Emerson2005}
\BIBentryALTinterwordspacing
J.~Emerson, R.~Alicki, and K.~{\.{Z}}yczkowski, ``Scalable noise estimation
  with random unitary operators,'' \emph{Journal of Optics B: Quantum and
  Semiclassical Optics}, vol.~7, no.~10, pp. S347--S352, Sep. 2005. [Online].
  Available: \url{https://doi.org/10.1088/1464-4266/7/10/021}
\BIBentrySTDinterwordspacing

\bibitem{Danker2009}
\BIBentryALTinterwordspacing
C.~Dankert, R.~Cleve, J.~Emerson, and E.~Livine, ``Exact and approximate
  unitary 2-designs and their application to fidelity estimation,'' \emph{Phys.
  Rev. A}, vol.~80, p. 012304, Jul 2009. [Online]. Available:
  \url{https://link.aps.org/doi/10.1103/PhysRevA.80.012304}
\BIBentrySTDinterwordspacing

\bibitem{quantumvolume}
\BIBentryALTinterwordspacing
A.~W. Cross, L.~S. Bishop, S.~Sheldon, P.~D. Nation, and J.~M. Gambetta,
  ``Validating quantum computers using randomized model circuits,'' \emph{Phys.
  Rev. A}, vol. 100, p. 032328, Sep 2019. [Online]. Available:
  \url{https://link.aps.org/doi/10.1103/PhysRevA.100.032328}
\BIBentrySTDinterwordspacing

\bibitem{CLOPS}
A.~Wack, H.~Paik, A.~Javadi-Abhari, P.~Jurcevic, I.~Faro, J.~M. Gambetta, and
  B.~R. Johnson, ``Quality, speed, and scale: three key attributes to measure
  the performance of near-term quantum computers,'' \emph{arXiv preprint
  arXiv:2110.14108v2}, 2021.

\bibitem{Lubinski:2021_arxiv}
\BIBentryALTinterwordspacing
T.~Lubinski, S.~Johri, P.~Varosy, J.~Coleman, L.~Zhao, J.~Necaise, C.~H.
  Baldwin, K.~Mayer, and T.~Proctor, ``{Application-Oriented Performance
  Benchmarks for Quantum Computing},'' 2021. [Online]. Available:
  \url{https://arxiv.org/abs/2110.03137}
\BIBentrySTDinterwordspacing

\bibitem{tomesh2022supermarq}
T.~Tomesh, P.~Gokhale, V.~Omole, G.~S. Ravi, K.~N. Smith, J.~Viszlai, X.-C. Wu,
  N.~Hardavellas, M.~R. Martonosi, and F.~T. Chong, ``Supermarq: A scalable
  quantum benchmark suite,'' 2022.

\bibitem{mesman2022qpack}
K.~Mesman, Z.~Al-Ars, and M.~Möller, ``Qpack: Quantum approximate optimization
  algorithms as universal benchmark for quantum computers,'' 2022.

\bibitem{Qscore}
\BIBentryALTinterwordspacing
S.~Martiel, T.~Ayral, and C.~Allouche, ``Benchmarking quantum coprocessors in
  an application-centric, hardware-agnostic, and scalable way,'' \emph{IEEE
  Transactions on Quantum Engineering}, vol.~2, p. 1–11, 2021. [Online].
  Available: \url{http://dx.doi.org/10.1109/TQE.2021.3090207}
\BIBentrySTDinterwordspacing

\bibitem{Dong_2021}
\BIBentryALTinterwordspacing
Y.~Dong and L.~Lin, ``Random circuit block-encoded matrix and a proposal of
  quantum linpack benchmark,'' \emph{Physical Review A}, vol. 103, no.~6, Jun.
  2021. [Online]. Available:
  \url{http://dx.doi.org/10.1103/PhysRevA.103.062412}
\BIBentrySTDinterwordspacing

\bibitem{Aharonov:2003}
\BIBentryALTinterwordspacing
D.~Aharonov, ``A simple proof that {T}offoli and {H}adamard are quantum
  universal,'' 2003. [Online]. Available:
  \url{https://arxiv.org/abs/quant-ph/0301040}
\BIBentrySTDinterwordspacing

\bibitem{Chow:2012}
\BIBentryALTinterwordspacing
J.~M. Chow, J.~M. Gambetta, A.~D. C\'orcoles, S.~T. Merkel, J.~A. Smolin,
  C.~Rigetti, S.~Poletto, G.~A. Keefe, M.~B. Rothwell, J.~R. Rozen, M.~B.
  Ketchen, and M.~Steffen, ``Universal quantum gate set approaching
  fault-tolerant thresholds with superconducting qubits,'' \emph{Phys. Rev.
  Lett.}, vol. 109, p. 060501, Aug 2012. [Online]. Available:
  \url{https://link.aps.org/doi/10.1103/PhysRevLett.109.060501}
\BIBentrySTDinterwordspacing

\bibitem{Schoot2022EvaluatingQscore}
W.~van~der Schoot, D.~Leermakers, R.~Wezeman, N.~Neumann, and F.~Phillipson,
  ``Evaluating the q-score of quantum annealers,'' in \emph{2022 IEEE
  International Conference on Quantum Software (QSW)}, 2022, pp. 9--16.

\bibitem{FGGS:2000}
E.~Farhi, J.~Goldstone, S.~Gutmann, and M.~Sipser, ``Quantum computation by
  adiabatic evolution,'' 2000.

\bibitem{Aharonov:2007}
\BIBentryALTinterwordspacing
D.~Aharonov, W.~van Dam, J.~Kempe, Z.~Landau, S.~Lloyd, and O.~Regev,
  ``Adiabatic quantum computation is equivalent to standard quantum
  computation,'' \emph{{SIAM} Journal on Computing}, vol.~37, no.~1, pp.
  166--194, Jan. 2007. [Online]. Available:
  \url{https://doi.org/10.1137/s0097539705447323}
\BIBentrySTDinterwordspacing

\bibitem{KN:1998}
\BIBentryALTinterwordspacing
T.~Kadowaki and H.~Nishimori, ``Quantum annealing in the transverse {I}sing
  model,'' \emph{Phys. Rev. E}, vol.~58, pp. 5355--5363, Nov 1998. [Online].
  Available: \url{https://link.aps.org/doi/10.1103/PhysRevE.58.5355}
\BIBentrySTDinterwordspacing

\bibitem{Lucas:2014}
\BIBentryALTinterwordspacing
A.~Lucas, ``{I}sing formulations of many {NP} problems,'' \emph{Frontiers in
  Physics}, vol.~2, 2014. [Online]. Available:
  \url{https://doi.org/10.3389/fphy.2014.00005}
\BIBentrySTDinterwordspacing

\bibitem{Kitaev1997}
\BIBentryALTinterwordspacing
A.~Y. Kitaev, ``Quantum computations: algorithms and error correction,''
  \emph{Russian Mathematical Surveys}, vol.~52, no.~6, pp. 1191--1249, Dec.
  1997. [Online]. Available:
  \url{https://doi.org/10.1070/rm1997v052n06abeh002155}
\BIBentrySTDinterwordspacing

\bibitem{BCG}
M.~Langione, J.-F. Bobier, L.~Krayer, H.~Park, and A.~Kumar, ``The race to
  quantum advantage depends on benchmarking,'' Boston Consulting Group, Tech.
  Rep., Feb 2022.

\bibitem{erdos1959random}
\BIBentryALTinterwordspacing
P.~Erdös and A.~Rényi, ``On random graphs,'' \emph{Publ. Math. Debrecen},
  vol.~6, pp. 290--297, 1959. [Online]. Available:
  \url{https://api.semanticscholar.org/CorpusID:253789267}
\BIBentrySTDinterwordspacing

\bibitem{erdos1960evolution}
\BIBentryALTinterwordspacing
P.~Erd{\H{o}}s and A.~R{\'e}nyi, ``On the evolution of random graphs,''
  \emph{Publ. Math. Inst. Hung. Acad. Sci}, vol.~5, no.~1, pp. 17--60, 1960.
  [Online]. Available: \url{https://doi.org/10.1515/9781400841356.38}
\BIBentrySTDinterwordspacing

\bibitem{FGG:2014}
E.~Farhi, J.~Goldstone, and S.~Gutmann, ``A quantum approximate optimization
  algorithm,'' 2014.

\bibitem{QUBO_problems}
\BIBentryALTinterwordspacing
F.~W. Glover and G.~A. Kochenberger, ``A tutorial on formulating {QUBO}
  models,'' \emph{CoRR}, vol. abs/1811.11538, 2018. [Online]. Available:
  \url{http://arxiv.org/abs/1811.11538}
\BIBentrySTDinterwordspacing

\bibitem{matula1976largest}
D.~W. Matula, ``The largest clique size in a random graph,'' Department of
  Computer Science, Southern Methodist University Dallas, Texas, Tech. Rep.,
  1976.

\bibitem{FahriGoldstoneGutmann:2014}
E.~Farhi, J.~Goldstone, and S.~Gutmann, ``A quantum approximate optimization
  algorithm,'' 2014.

\bibitem{runtime}
I.~Quantum, ``Qiskit runtime,''
  \url{https://quantum-computing.ibm.com/lab/docs/iql/runtime/}.

\bibitem{Weidenfeller2022}
J.~Weidenfeller, L.~C. Valor, J.~Gacon, C.~Tornow, L.~Bello, S.~Woerner, and
  D.~J. Egger, ``Scaling of the quantum approximate optimization algorithm on
  superconducting qubit based hardware,'' \emph{Quantum}, vol.~6, p. 870, dec
  2022.

\bibitem{BLEKOS20241}
K.~Blekos, D.~Brand, A.~Ceschini, C.-H. Chou, R.-H. Li, K.~Pandya, and
  A.~Summer, ``A review on quantum approximate optimization algorithm and its
  variants,'' \emph{Physics Reports}, vol. 1068, pp. 1--66, 2024, a review on
  Quantum Approximate Optimization Algorithm and its variants.

\bibitem{github}
R.~Wezeman, W.~van~der Schoot, and E.~Aguilera, ``{TNO Quantum / qscore},''
  \url{https://github.com/TNO-Quantum/qscore}.

\bibitem{guadalupe16}
IBM, ``ibmq\_guadalupe,''
  \url{https://quantum-computing.ibm.com/services/resources?tab=systems&amp;view=grid&amp;system=ibmq_guadalupe}.

\bibitem{starmon5}
QuTech, ``Quantum inspire starmon-5 fact sheet,''
  \url{https://www.quantum-inspire.com/backends/starmon-5/}, 6 2020.

\bibitem{AA:2013}
\BIBentryALTinterwordspacing
S.~Aaronson and A.~Arkhipov, ``The computational complexity of linear optics,''
  \emph{Theory of Computing}, vol.~9, no.~4, pp. 143--252, 2013. [Online].
  Available: \url{https://theoryofcomputing.org/articles/v009a004}
\BIBentrySTDinterwordspacing

\bibitem{Arrazola_2018}
J.~M. Arrazola and T.~R. Bromley, ``Using {G}aussian boson sampling to find
  dense subgraphs,'' \emph{Physical Review Letters}, vol. 121, no.~3, Jul 2018.

\bibitem{Banchi2020}
\BIBentryALTinterwordspacing
L.~Banchi, M.~Fingerhuth, T.~Babej, C.~Ing, and J.~M. Arrazola, ``Molecular
  docking with {G}aussian boson sampling,'' \emph{Science Advances}, vol.~6,
  no.~23, p. eaax1950, 2020. [Online]. Available:
  \url{https://www.science.org/doi/abs/10.1126/sciadv.aax1950}
\BIBentrySTDinterwordspacing

\bibitem{Haverly2021}
A.~R. Haverly, ``A comparison of quantum algorithms for the maximum clique
  problem,'' Master's thesis, Rochester Institute of Technology., 5 2021.

\bibitem{quandela}
Quandela, ``Quandela, photonic quantum computers,''
  \url{https://www.quandela.com/}.

\bibitem{photonic_sampler}
Xanadu, ``sf.apps.sample,''
  \url{https://strawberryfields.readthedocs.io/en/stable/code/api/strawberryfields.apps.sample.html}.

\bibitem{Glover1989}
\BIBentryALTinterwordspacing
F.~Glover, ``Tabu search{\textemdash}part {I},'' \emph{{ORSA} Journal on
  Computing}, vol.~1, no.~3, pp. 190--206, Aug. 1989. [Online]. Available:
  \url{https://doi.org/10.1287/ijoc.1.3.190}
\BIBentrySTDinterwordspacing

\bibitem{Glover1990}
\BIBentryALTinterwordspacing
------, ``Tabu search{\textemdash}part {II},'' \emph{{ORSA} Journal on
  Computing}, vol.~2, no.~1, pp. 4--32, Feb. 1990. [Online]. Available:
  \url{https://doi.org/10.1287/ijoc.2.1.4}
\BIBentrySTDinterwordspacing

\bibitem{Pelofske}
\BIBentryALTinterwordspacing
E.~Pelofske, G.~Hahn, and H.~Djidjev, ``Solving large maximum clique problems
  on a quantum annealer,'' in \emph{Quantum Technology and Optimization
  Problems}.\hskip 1em plus 0.5em minus 0.4em\relax Springer International
  Publishing, 2019, pp. 123--135. [Online]. Available:
  \url{https://doi.org/10.1007%2F978-3-030-14082-3_11}
\BIBentrySTDinterwordspacing

\bibitem{barbaresco2024bacq}
F.~Barbaresco, L.~Rioux, C.~Labreuche, M.~Nowak, N.~Olivier, D.~Nicolazic,
  O.~Hess, A.-L. Guilmin, R.~Wang, T.~Sassolas, S.~Louise, K.~Snizhko,
  G.~Misguich, A.~Auffèves, R.~Whitney, E.~Vergnaud, and F.~Schopfer, ``Bacq
  -- application-oriented benchmarks for quantum computing,'' 2024.

\bibitem{haverly2021grov}
A.~Haverly and S.~López, ``Implementation of {G}rover’s algorithm to solve
  the maximum clique problem,'' in \emph{2021 IEEE Computer Society Annual
  Symposium on VLSI (ISVLSI)}, 2021, pp. 441--446.

\bibitem{Tilly_2022}
\BIBentryALTinterwordspacing
J.~Tilly, H.~Chen, S.~Cao, D.~Picozzi, K.~Setia, Y.~Li, E.~Grant, L.~Wossnig,
  I.~Rungger, G.~H. Booth, and J.~Tennyson, ``The variational quantum
  eigensolver: A review of methods and best practices,'' \emph{Physics
  Reports}, vol. 986, p. 1–128, Nov. 2022. [Online]. Available:
  \url{http://dx.doi.org/10.1016/j.physrep.2022.08.003}
\BIBentrySTDinterwordspacing

\end{thebibliography}

\end{document}